\documentclass[showpacs,twocolumn,aps]{revtex4}
\usepackage{amsmath}
\usepackage{amsfonts}
\usepackage{amssymb}
\usepackage{graphicx}
\usepackage{color}
\usepackage{epsf}
\begin{document}
\title{Velocity Distribution in a Viscous Granular Gas}
\author{Alexandre Rosas$^{1,2}$}
\author{Daniel ben-Avraham$^3$}
\author{Katja Lindenberg$^{1,4}$}
\affiliation{${}^1$Department of Chemistry and Biochemistry, University of
California San Diego, La Jolla, CA 92093-0340\\
${}^2$Departamento de F\'{\i}sica, Universidade Federal da Para\'{\i}ba, 
Jo\~ao Pessoa, PB 58.059-970, Brazil\\
${}^3$Physics Department, Clarkson University, Potsdam NY
13699-5820\\
${}^4$Institute for Nonlinear Science, University of
California San Diego, La Jolla, CA 92093-0340}
\begin{abstract}
We investigate the velocity relaxation of a viscous one-dimensional
granular gas, that is, one in which neither energy nor momentum is
conserved in a collision.  Of interest is the distribution of velocities
in the gas as it cools, and the time dependence of the relaxation
behavior.  A Boltzmann equation of instantaneous binary collisions 
leads to a two-peaked distribution with each peak relaxing to zero
velocity as $1/t$ while each peak also narrows as $1/t$. 
Numerical simulations of grains on a line also lead to a
double-peaked distribution that narrows as $1/t$.
A Maxwell approximation leads to a single-peaked distribution
about zero velocity with power-law wings.  This distribution
narrows exponentially.  In either case, the relaxing distribution is not
of Maxwell-Boltzmann form.

\end{abstract}
%45.70.-n Granular Systems
%05.20.Dd Kinetic Theory
%05.70.Ln Nonequilibrium and irreversible thermodynamics 
%83.10.Rs Computer simulation of molecular and particle dynamics

\maketitle

\section{Introduction}
\label{introduction}
Velocity distributions in dilute granular gases are generically away from
equilibrium because the collision processes in such gases are dissipative.
Even in the dilute limit the velocities of different
particles may be strongly correlated~\cite{moon}, and therefore the
usual description in terms of single particle distribution functions may
not be sufficient to determine all the properties of the granular gas. 
Nevertheless, the single particle distribution $P(r,v,t)$ contains
important information, and is particularly interesting because in a
granular gas it typically deviates from the Maxwell-Boltzmann distribution.  
Quite aside from possible spatial inhomogeneity effects such as particle
clustering, even the single
particle velocity distribution $P(v,t)$ in general differs from the
Maxwell-Boltzmann form.   The deviations of these single particle
distributions from
the usual ``universal'' behavior in ordinary gases has been a subject of
intense interest in recent years~\cite{warr,olafsen,losert,kudrolli,rouyer}.
These deviations can be observed in granular gases that achieve a steady
state because external forcing balances the dissipative collisions among
particles~\cite{grossman,barrat,swinney}, or they can be observed in
unforced gases as they cool down~\cite{young1,plus,bmp,kb,eb,bc}.

Typically, gases equilibrate or achieve a steady state via collision
processes.  Conservation of energy and momentum imply  Maxwellian
velocity statistics in three dimensions~\cite{jcm}.  In lower dimensions
the situation is more complicated; for example, a gas of hard spheres in
one dimension simply retains its initial distribution forever.  
In granular media the situation is in any case more complex because energy is
certainly not conserved~\cite{pl}. Most work on relaxation in granular
materials, in any dimension, has focused on the interesting consequences
of energy non-conservation. 
A particularly interesting behavior of granular gases induced by energy
non-conservation is the so-called ``inelastic
collapse,''  whereby the energy of the gas goes to zero in a finite
time~\cite{young1}.  In the ``quasi-elastic limit'' the inelastic
collapse is avoided, and one obtains non-trivial asymptotic
velocity distributions~\cite{grossman,barrat,plus} with features similar
to the ones obtained in this paper.

In the presence of friction, not only is
energy not
conserved, but momentum is not conserved either, leading to 
interesting new relaxation behavior~\cite{swinney,our1,our2,our3}.  
In fact, we have
recently shown~\cite{our3} that in the absence of conservation laws
random linear mixing can lead to velocity distributions with
algebraic or exponential tails, with nontrivial characteristic exponents.  
In general, conservation laws play a crucial role in the universality of
the usual velocity distribution properties.
  
Our focus here is the effect of viscosity, and consequently,
of momentum non-conservation, on the velocity distribution as a
one-dimensional dilute granular gas cools down.
The model consists of $N$ grains on a line (or a
circle, since we will use periodic boundary conditions).
The grains move freely except during collisions,
governed by the Hertz potential,
\begin{equation}
\begin{array}{l l l}
V(\delta_{k,k+1})&=\frac{\displaystyle a}{\displaystyle n}|\delta|^{n}_{k,k+1}, \qquad &\delta\leq 0,\\ \\
V(\delta_{k,k+1})&= 0, \qquad &\delta >0.
\end{array}
\label{eq:hertz}
\end{equation}
Here $\delta_{k,k+1} \equiv y_{k+1} - y_{k}$,
$y_k$ is the displacement of granule $k$ from its equilibrium position,
and $a$ is a prefactor determined by Young's modulus and Poisson's ratio. 
The exponent $n$ is $5/2$ for spheres, it is $2$ for cylinders, and
in general depends on geometry.  In this paper we only consider cylindrical
grains, which leads to considerable simplification while still capturing 
important general features of the non-Maxwellian distributions.  We stress
that the one-sided (only repulsion) granular potential even with
$n=2$ is entirely different from a two-sided (repulsion and attraction)
harmonic potential.

The displacement of the $k$-th granule ($k=1,2, \ldots, N$) of mass $m$
in the chain
from its equilibrium position in a frictional medium
is governed by the equation of motion
%\begin{equation}
%\begin{split}
\begin{eqnarray}
m \frac{\mathrm{d}^2{y}_k}{\mathrm{d} \tau^2} &=& -\tilde{\gamma}
\frac{\mathrm{d}{y}_k}{\mathrm{d}\tau} \nonumber \\
&&- a (y_k - y_{k+1})^{n-1}
\theta (y_k - y_{k+1}) \nonumber\\
&&+ a (y_{k-1} - y_{k})^{n-1}
\theta (y_{k-1} - y_{k}).
\label{eq:motion}
%\end{split}
%\end{equation}
\end{eqnarray}
Here $\tilde{\gamma}$ is the friction coefficient and
$\theta(y)$ is the Heaviside function, $\theta(y)=1$ for $y>0$, 
$\theta(y)=0$ for $y<0$, and $\theta(0)=1/2$. 
The Heaviside function ensures that two particles interact only when
in contact, that is, only when the particles are loaded. Note
that for periodic boundary conditions $y_{k+N}=y_k$.
In terms of the rescaled variables and parameters
\begin{equation}
%y_k = \left( \frac{m v_0^2}{a} \right) ^{1/n} x_k, &\qquad&
y_k = C_n  x_k, \quad
\tau = \frac{1}{v_0} C_n t, \quad
\gamma =\frac{\tilde{\gamma}}{m v_0} C_n, 
\end{equation}
where $C_n\equiv (mv_0^2/a)^{1/n}$, Eq.~(\ref{eq:motion}) can
be rewritten as
%\begin{equation}
\begin{eqnarray}
\ddot{x}_k &=& -\gamma \dot{x}_k - (x_k - x_{k+1})^{n-1} \theta (x_k -
x_{k+1}) \nonumber\\
&&+  (x_{k-1} - x_{k})^{n-1} \theta (x_{k-1} -
x_{k}).\label{eq:motion_rescaled}
%\end{equation}
\end{eqnarray}
A dot denotes a derivative with respect to $t$. The velocity $v_0$ is
an arbitrary choice in terms of which other velocities are now expressed, i.e.,
it settles the energy scale of the system.

Although the problem might appear relatively simple because it
is one-dimensional
and quasi-linear, the one-sidedness of the potential leads to
analytic complexities even in the dissipationless
case~\cite{nesterenko,hinch,our2}, and even greater complexities in the
presence of dissipation~\cite{our1,our2}. Our purpose in this paper is to
explore analytic approximations appropriate for \emph{low densities}.
The low density feature of these approximations is implemented via the
assumption that the collisions are always \emph{binary}, that is, that only
two granules at a time are members of any collision event, and that at any
moment of time there is at most one collision.  

We explore two low-density analytic approximations. One is based on the
Boltzmann equation for binary collisions, presented in
Sec.~\ref{sec:defs} and elaborated in Sec.~\ref{sec:Boltzmann},
and the other is an even simpler model 
that we call the Maxwell model, presented in Sec.~\ref{sec:Maxwell}.
The detailed description of a
binary collision that must be used as an input in either of these models
is also presented in Sec.~\ref{sec:defs}.  To assess the success of
each analytic model, a comparison of the approximate model
results with numerical simulations of the entire chain
is carried out in Sec.~\ref{sec:numerical}. In that section we also
summarize the outcome of our analysis.

\section{Boltzmann Equation and Collisions}
\label{sec:defs}
Our analytic starting point in the following sections
is the Boltzmann equation for binary collisions
in a spatially uniform gas, which
describes the rate of change of the probability distribution of velocities,
$P(v,t)$.
If $u_1$ and $u_2$ are the initial velocities of a pair of
particles just before a collision and
and $u_1^\prime$ and $u_2^\prime$ their velocities just after,
then the Boltzmann equation is
\begin{eqnarray}
\frac{\partial}{\partial t} P(v,t) = \iint du_1\; du_2
|u_1-u_2|P(u_1,t)P(u_2,t)\nonumber\\ \nonumber\\
%&\times&\left[ \delta(v-u_1^\prime) + \delta(v-u_2^\prime)
\times\left[ \delta(v-u_1^\prime) + \delta(v-u_2^\prime)
-\delta(v-u_1) - \delta(v-u_2)\right].
\label{Beq}
\end{eqnarray}
Since the problem is one-dimensional, one can keep track of the precise
conditions under which a collision between two particles of given
velocities will or will not occur, and how these events will change the
distribution function. 

The binary collision approximation implies that at any one time there
is at most one pair of loaded grains, so we need to consider in detail the
fate of a colliding pair from the beginning to the end of a
collision~\cite{our1}.
At the moment of the start of the collision, which we
call $t=0$, the velocities of the grains are $u_1$ and $u_2$. 
The collision ends at the time $\tau=2\pi/\sqrt{8-\gamma^2}$, when the
grains lose contact.  
It is important to notice that for $ n=2 $ the collision time 
\emph{is independent} of the initial condition.  This is
a feature that makes the cylindrical granule geometry much simpler than
other shapes.
Without loss of generality we take $u_1>u_2 $, since we can
always relabel particles to make it so.  The velocities 
at the moment of separation $\tau$ are found to be~\cite{our1}
\begin{equation}
\begin{split}
u_1^\prime &=  u_1 \frac{\mu^2 - \mu}{2} +
u_2 \frac{\mu^2 + \mu}{2},\\
u_2^\prime &=  u_1 \frac{\mu^2 + \mu}{2} +
u_2 \frac{\mu^2 - \mu}{2},\label{eq:vel}
\end{split}
\end{equation}
where $\mu \equiv  e^{-\gamma \tau/2}$.  

In our further analysis we think of the distance traveled by the
granules during a collision as negligible.  For small damping $\gamma$ this
distance is $(u_1+u_2) \pi/\sqrt{8}$,
which is to be compared to the mean distance between particles.
The latter can be made arbitrarily large by lowering the density.
We also take the collision time between two granules as instantaneous.
This collision time for small damping
is $\tau \approx \pi/\sqrt{2}$, to be compared with the
typical mean free time of flight of a particle between collisions.
With these approximations, the only role played by the viscosity is
dictated by the collision rule~(\ref{eq:vel}).  
These assumptions might conceivably be problematic for the most
energetic particles that may travel a relatively long distance during
a collision and a relatively short distance between collisions, but
explicit analysis of these extreme events is difficult and probably not
important at sufficiently low densities. 

\section{Boltzmann Problem}
\label{sec:Boltzmann}

We now start with the Boltzmann equation (\ref{Beq}) in which the
collision rate depends on the difference in the velocities of the
colliding pair.
To write the explicit Boltzmann equation for our system, we must collect
all the instances that lead to a collision, that is, all the instances for
which one or another of the $\delta$-functions in Eq.~(\ref{Beq}) can be
satisfied.  There are six possible events that contribute to a change in
$P(v,t)$.  We list the contribution to $\partial P/\partial t$ from
each event:

\begin{list}{}
\item {(a)} $(v, u)
\longrightarrow (u_1^\prime, u_2^\prime); \qquad v > u $\\
Implementing the third $\delta$-function in Eq.~(\ref{Beq}) we find
\begin{equation}
\left[\frac{\partial}{\partial t} P(v,t)\right]_{(a)} =
-P(v,t)\int_{-\infty}^v du (v-u) P(u,t).
\label{eqa}
\end{equation}
\vskip 5pt
\item{(b)}
$(u, v) \longrightarrow (u_1^\prime, u_2^\prime);\qquad u > v $\\
Implementation of the fourth $\delta$-function in Eq.~(\ref{Beq}) leads
to
\begin{equation}
\left[\frac{\partial}{\partial t} P(v,t)\right]_{(b)} =
-P(v,t)\int_v^\infty du (u-v) P(u,t).
\label{eqb}
\end{equation}
\vskip 5pt
\item{(c)}
$(u_1, u_2) \longrightarrow (v, u_2^\prime); \qquad u_1 > u_2 $\\
Here we implement the first $\delta$-function in Eq.~(\ref{Beq}) to find
\begin{eqnarray}
\left [\frac{\partial }{\partial t}P(v,t)\right]_{(c)}
= \int_{-\infty}^{v/\mu^2} \left(\frac{2v}{\mu^2 - \mu} - \frac{2
\mu^2u}{\mu^2 - \mu} \right)\nonumber\\
\times P\left(\frac{2v}{\mu^2 - \mu} -
\frac{(\mu^2
+ \mu)u}{\mu^2 - \mu} , t\right) P(u, t) du.
\label{eqc}
\end{eqnarray}
\item {(d)}
$(u_1, u_2) \longrightarrow (v, u_2^\prime); \qquad u_1<u_2 $\\
We again implement the first $\delta$-function in Eq.~(\ref{Beq}):
\begin{eqnarray}
\left [\frac{\partial }{\partial t}P(v,t)\right]_{(d)}
= \int_{v/\mu^2}^\infty
\left(\frac{2\mu^2 u}{\mu^2 - \mu} -
\frac{2 v}{\mu^2 - \mu} \right)\nonumber\\
\times P\left(\frac{2v}{\mu^2 - \mu} -
\frac{(\mu^2
+ \mu)u}{\mu^2 - \mu} , t\right) P(u, t) du.
\label{eqd}
\end{eqnarray}
\item {(e)}
$(u_1,u_2) \longrightarrow (u_1^\prime, v); \qquad u_1>u_2$\\
Now we implement the second $\delta$-function in Eq.~(\ref{Beq}):
\begin{eqnarray}
\left [\frac{\partial }{\partial t}P(v,t)\right]_{(e)}
= \int_{v/\mu^2}^\infty
\left(\frac{2\mu^2 u}{\mu^2 + \mu}
- \frac{2 v}{\mu^2 + \mu} \right)\nonumber\\
\times P\left(\frac{2v}{\mu^2 + \mu} -
\frac{(\mu^2
- \mu)u}{\mu^2 + \mu} , t\right) P(u, t) du.
\label{eqe}
\end{eqnarray}
\item {(f})
$(u_1,u_2) \longrightarrow (u_1^\prime,v);  \qquad u_1<u_2$\\
Again we implement the second $\delta$-function in Eq.~(\ref{Beq}):
\begin{eqnarray}
\left [\frac{\partial }{\partial t}P(v,t)\right]_{(f)}
= \int_{-\infty}^{v/\mu^2}
\left(\frac{2v}{\mu^2 + \mu}
- \frac{2 \mu^2 u}{\mu^2 + \mu} \right)\nonumber\\
\times P\left(\frac{2v}{\mu^2 + \mu} -
\frac{(\mu^2
- \mu)u}{\mu^2 + \mu} , t\right) P(u, t) du.
\label{eqf}
\end{eqnarray}
\end{list} 
The full equation for $\partial P(v,t)/\partial t$ is the sum of these
six contributions.
While the collisions of types~(a) and~(b) decrease the probability
density $ P(v, t), $ collisions~(c) to~(f) increase it.

We assume a scaling solution of the form, 
\begin{equation}
P(v, t) = \frac{1}{\phi (t)} F\left (\frac{v}{\phi (t)} \right ),
\label{eq:scaling}
\end{equation} 
and define 
\begin{equation}
x \equiv \frac{v}{\phi(t)}, \qquad y \equiv \frac{u}{\phi(t)}.
\end{equation}
The Boltzmann equation then is 
\begin{eqnarray}
%\begin{split}
\frac{\partial P(v, t)}{\partial t} &&=
-\frac{\dot{\phi} (t)}{\phi^2 (t)} \left ( F(x)
+ x F^\prime(x) \right )
\nonumber\\
&&= - \int_{-\infty}^x (x - y) F(x) F(y) dy
\nonumber\\
&&- \int_x^{\infty} (y - x) F(x) F(y) dy\nonumber\\
&& +  \int_{-\infty}^{x/\mu^2} \left(\frac{2x}{\mu^2 - \mu} 
- \frac{2 \mu^2y}{\mu^2 - \mu} \right)\nonumber\\
&&\times
F\left(\frac{2x}{\mu^2 - \mu} - \frac{(\mu^2 + \mu)y}{\mu^2 -
\mu} \right) F(y) dy \nonumber\\
&& + \int_{x/\mu^2}^{\infty} \left(\frac{2 \mu^2y}{\mu^2 - \mu}
 - \frac{2x}{\mu^2 - \mu}\right) \nonumber\\
&&\times
F\left(\frac{2x}{\mu^2 - \mu} - \frac{(\mu^2 + \mu)y}{\mu^2 -
\mu} \right) F(y) dy\nonumber \\
&& + \int_{x/\mu^2}^{\infty} \left(\frac{2 \mu^2y}{\mu^2 + \mu}
 - \frac{2x}{\mu + \mu^2}  \right) \nonumber\\
&&\times
F\left(\frac{2x}{\mu^2 + \mu} - \frac{(\mu^2 - \mu)y}{\mu^2 +
\mu} \right) F(y) dy \nonumber\\
&&+ \int_{-\infty}^{x/\mu^2} \left(\frac{2x}{\mu^2 + \mu}
  - \frac{2\mu^2y}{\mu + \mu^2} \right) \nonumber\\
&&\times
F\left(\frac{2x}{\mu^2 + \mu} - \frac{(\mu^2 - \mu)y}{\mu^2 +
\mu}\right) F(y) dy.
%\end{split}
\end{eqnarray} 

Defining $ z= \frac{2x}{\mu^2 - \mu}  -
\frac{(\mu^2 + \mu)y}{\mu^2 - \mu} $ and recognizing that the
dummy variable $z$ can be relabeled as $y$ after the transformation,
one finds that the third term on the right hand side of this equation
becomes proportional to the fourth, and the fifth to the sixth, so that
these pairs can be combined to yield the simpler expression
%\begin{equation}
%\begin{split}
\begin{eqnarray}
 -\frac{\dot{\phi} (t)}{\phi^2 (t)}&& \left ( F(x) + x F^\prime(x)
\right )\nonumber\\
&&= - \int_{-\infty}^x (x - y) F(x) F(y) dy \nonumber\\
&&- \int_x^{\infty} (y - x) F(x) F(y) dy\nonumber\\
&& + \frac{2}{\mu + 1} \int_{-\infty}^{x/\mu^2} \left(\frac{2x}
{\mu^2 + \mu}  - \frac{2 \mu^2y}{\mu^2 + \mu}\right)\nonumber\\
&&\times
F\left(\frac{2x}{\mu^2 + \mu} - \frac{(\mu^2 - \mu)y}
{\mu^2 + \mu}\right) F(y) dy \nonumber\\
&& + \frac{2}{\mu + 1} \int_{x/\mu^2}^{\infty} \left(
\frac{2 \mu^2y}{\mu^2 + \mu} - \frac{2x}{\mu^2 
+ \mu} \right)\nonumber\\
&&\times
F\left(\frac{2x}{\mu^2 + \mu} -
\frac{(\mu^2 - \mu)y}{\mu^2 + \mu} \right) F(y) dy.
%\end{split}
\label{eq:Breduced}
%\end{equation} 
\end{eqnarray}

Since the only explicit time dependence resides in the
$\frac{\dot{\phi} (t)}{\phi^2 (t)}$ term, this term must be
constant. Integrating this term leads to $ \phi (t) = \frac{\phi(0)}{1+c
t}$, where $c$ is a constant. Therefore, asymptotically,
\begin{equation}
\phi (t) \sim t^{-1}.
\label{eq:Bscaling}
\end{equation}

We have not found an analytic solution to Eq.~(\ref{eq:Breduced}). We
therefore simulate the equation numerically (next subsection) and
implement a further approximation that leads to an analytic solution that
we can compare with the numerical results (subsequent subsection).

\subsection{Simulation of the Boltzmann equation}
\label{sec:Bsimul}

We directly simulate the Boltzmann equation using the following algorithm:
\begin{enumerate}
\item We start with $ N $ grains whose velocities are independently
assigned accordingly to an initial distribution $P(v, 0)$.\label{it:algoa}
\item We choose one pair of grains with probability proportional to
the modulus of their relative velocity and let them collide, using the
collision rule~\ref{eq:vel}.
\item Time is incremented by twice the inverse of the modulus of the
pre-collisional relative velocity.\label{it:algoc}  The factor of 2
accounts for picked pairs that do not collide, since our algorithm
forces a collision at each step.
\item We iterate steps~\ref{it:algoa} to~\ref{it:algoc} many times and
for many samples.
\end{enumerate} 

\begin{figure}
\begin{center}
\includegraphics[width=8cm]{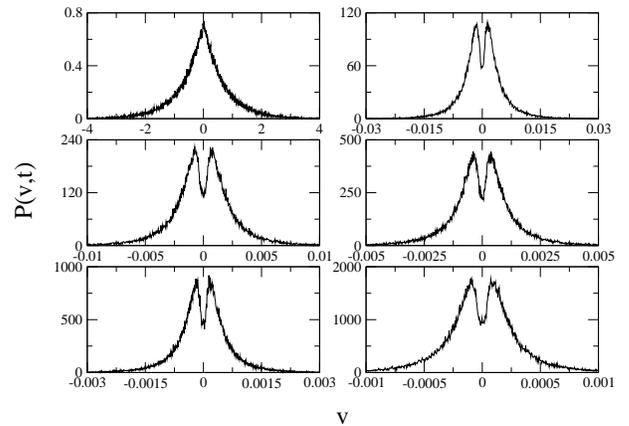}
\caption{Velocity distribution obtained using the simulation
algorithm detailed in
Sec.~\ref{sec:Bsimul}. From left to right and top to bottom,
the panels correspond to time 0, 8000, 16000, 32000, 64000 and
128000 in the adimensional units used in this paper. The initial
distribution is a symmetric exponential. Notice the change in the
scales as time proceeds.\label{fig:dist}}
\end{center}
\end{figure}

\begin{figure}
\begin{center}
\vskip 10pt
\includegraphics[width=8cm]{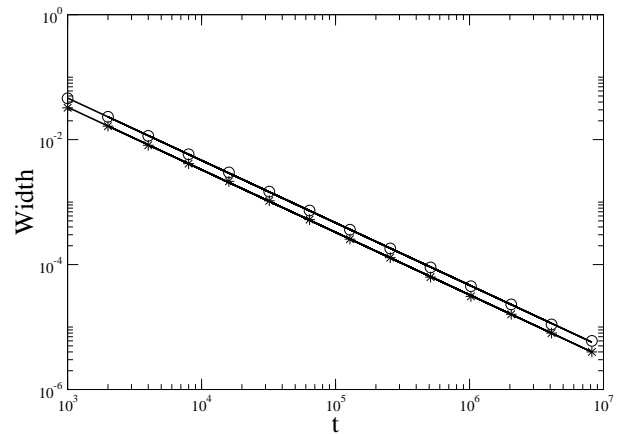}
\caption{Width of the velocity distribution (circles) and the 
width of each peak (stars). The lines represent the best fit
of the width vs time for the full distribution (exponent $-0.998$)
and for each peak (exponent $-1.001$).\label{fig:width}}
\end{center}
\end{figure}

\begin{figure}
\begin{center}
\includegraphics[width=8cm]{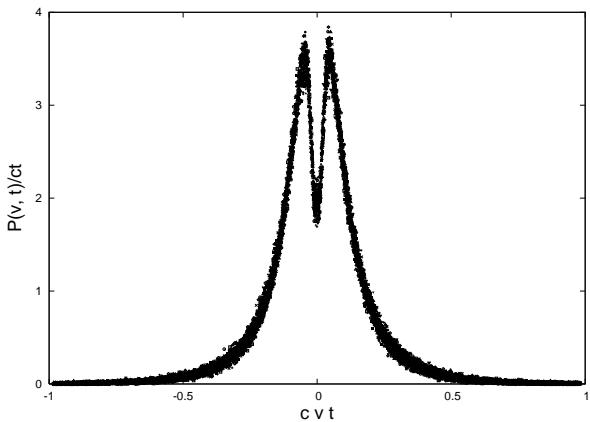}
\caption{Asymptotic behavior of the scaled velocity distribution. Different
symbols stand for different times
($1000\times 2^n$ with $n=0,1,\cdots,13$).
%(1000, 2000, 4000, 8000, 16000, 32000,
%64000, 128000, 256000, 512000, 1024000, 2048000,
%4096000 and 8192000).
The constant $c$ in the figure is 
arbitrary and was chosen to facilitate comparison with
Fig.~\ref{fig:scaling}. \label{fig:scaling-boltz}}
\end{center}
\end{figure}

In our simulations we took $N = 100$ and averaged our results over
1000 samples. In Fig.~\ref{fig:dist} we show the resulting velocity
distribution for $\gamma=0.9$ at different times. It is clear that the initial
symmetric and single peaked distribution develops
two distinct peaks as it starts to collapse to the ultimate
equilibrium distribution, a $\delta$-function at $v = 0 $. Moreover,
in Fig.~\ref{fig:width} we show that the width of the distribution
and the width of each peak both decay as $ 1/t $. 
In the simulation underlying this
figure the initial distribution was a double peaked Gaussian,
chosen because it converges quickly. We find the same asymptotic behavior
for the initially exponential distribution, namely, 
that each peak moves inward as $1/t$ and also shrinks as $1/t$.
The validity of the scaling solution Eq.~(\ref{eq:scaling}) is clearly
evident in the scaled rendition of the velocity distribution shown in 
Fig.~\ref{fig:scaling-boltz} for different times. 

\subsection{Two-particle model}
\begin{figure}
\begin{center}
\includegraphics[width=8cm]{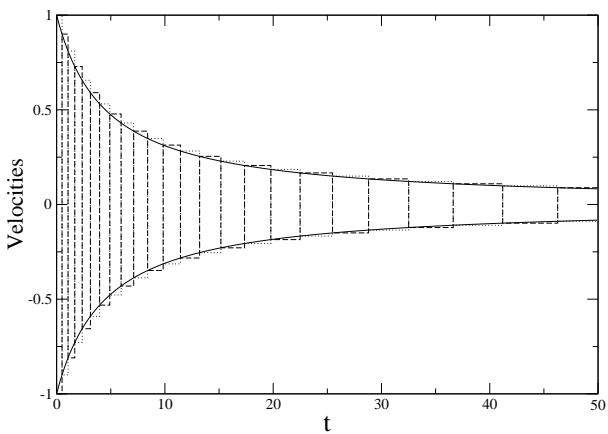}
\caption{Dashed lines and dotted lines are respectively velocities $u_n$
and $v_n$ on a ring with two particles, with $u_0=-v_0=1$ and $l=1$.
The solid curves are the envelopes $A(t)$ (upper) and $B(t)$ (lower).
\label{fig:velocities}}
\end{center}
\end{figure}

While the Boltzmann equation (\ref{eq:Breduced}) does not appear amenable
to analytic solution, we can formulate a simpler model that may
incorporate the main effects of the system.
In this model we have an ensemble of rings of size $l$ each containing
only two particles. In each ring, the dynamics is perfectly
deterministic: the two balls will keep colliding with each other,
back and forth, moving toward each other with ever decreasing velocities.
For each one ring, after $n$ collisions, the
velocities, which we call $u_n$ and $v_n $, are obtained by repeated
application of the collision rule (\ref{eq:vel}),
\begin{equation}
\begin{split}
u_n &= \frac{\mu^{2n}+(\textcolor{blue}{-}\mu)^n}{2} u_0 + \frac{\mu^{2n}-(\textcolor{blue}{-}\mu)^n}{2} v_0,\\
v_n &= \frac{\mu^{2n}-(\textcolor{blue}{-}\mu)^n}{2} u_0 + \frac{\mu^{2n}+(\textcolor{blue}{-}\mu)^n}{2} v_0,\label{eq:boltz-n}
\end{split}
\end{equation} 
where $u_0$ and $v_0$ are the initial velocities.
Note that each of these velocities alternates from positive to negative
as the particles move in one direction and then another in the ring,
as illustrated in Fig.~\ref{fig:velocities}.
The time that has elapsed by the $n^{th}$ collision is obtained as follows.
Since the grains are in a ring, the distance they have to travel
between two collisions is $l$. The travel time is $ l/|u-v|$, where $u$ and
$v$ are their current velocities between collisions. Hence,
\begin{equation}
t_n = \frac{l}{|u_0 - v_0|}\sum_{k=0}^{n-1} \mu^{-k}
= \frac{l}{|u_0 - v_0|} \frac{\frac{1}{\mu^n}-1}{\frac{1}{\mu}-1}
\label{eq:tn}
\end{equation} 

The sign alternation of the velocities in Eq.~(\ref{eq:boltz-n}) is not
important in the effort to understand the long time behavior of the
distribution.  We can approximate the
velocities by envelope functions $A(t)$ (upper curve)
and $B(t)$ (lower curve), as illustrated in
Fig.~\ref{fig:velocities}:
\begin{equation}
\begin{split}
u_n &= A(t) u_0 + B(t) v_0,\\
v_n &= B(t) u_0 + A(t) v_0.\label{eq:boltz-t}
\end{split}
\end{equation} 
The initial velocities $u_0$ and $v_0$ are arbitrary, and have only been
chosen of equal magnitude for purposes of illustration in the figure.
The envelope functions can be found by solving Eq.~(\ref{eq:tn}) for
$\mu^n$, substituting this into Eq.~(\ref{eq:boltz-n})
ignoring the minus signs in the $(-\mu)^n$ factors, and setting
$t_n\equiv t$.  One finds
\begin{equation}
\begin{split}
A(t) &= \frac{1}{2}\left[ 1 + \frac{|u_0 - v_0|}{l}\left
(\frac{1}{\mu} - 1\right ) t\right ]^{-2}\\
&+ \frac{1}{2}\left[ 1 +
\frac{|u_0 - v_0|}{l}\left (\frac{1}{\mu} - 1\right ) t\right
]^{-1},\\ \\
B(t) &= \frac{1}{2}\left[ 1 + \frac{|u_0 - v_0|}{l}\left (\frac{1}{\mu}
- 1\right ) t\right ]^{-2}\\
&- \frac{1}{2}\left[ 1 + \frac{|u_0 - v_0|}{l}\left (\frac{1}{\mu} - 1\right ) t\right ]^{-1}.
\end{split}
\end{equation} 
Hence, if initially the particles had a velocity distribution
$ P(v, 0), $ then the velocity distribution as a function of time is
\begin{eqnarray}
P(v, t) &=& \iint \delta (v - A(t) u_0 - B(t) v_0) \nonumber\\
\nonumber\\
&&\times P(u_0, 0) P(v_0, 0) du_0 dv_0.
\end{eqnarray} 

Asymptotically,
\begin{equation}
A(t) = - B(t) \sim \frac{1}{2} \frac{l}{|u_0 - v_0| t}
\left (\frac{1}{\mu} - 1\right )^{-1},
\end{equation}
so that
\begin{eqnarray}
P(v, t) &=& \iint \delta \left  (v -  \frac{1}{2}
\frac{l}{t}\left (\frac{1}{\mu} - 1\right )^{-1}
\mathrm{sgn} (u_0 -  v_0) \right )\nonumber\\
\nonumber\\
&&\times P(u_0, 0) P(v_0, 0) du_0 dv_0.
\end{eqnarray}
Therefore, $P(u,t)$ consists of two $\delta$-peaks,
one at positive and one at
negative velocities.  If the initial distribution is symmetric about
zero velocity and is properly normalized, then
\begin{eqnarray}
P(v,t)&\sim&
\frac{1}{2}\delta\left(v- \frac{l}{2t}\left (\frac{1}{\mu}
- 1\right )^{-1}\right)\nonumber\\
&&+
\frac{1}{2}\delta\left(v+ \frac{l}{2t}\left (\frac{1}{\mu}
- 1\right )^{-1}\right).
\end{eqnarray}
The peaks thus move toward the final velocity $u=0$ as $1/t$.
This time dependence is in agreement with the Boltzmann equation analysis.

The two $\delta$-peaks here reflect the fact that the magnitude of the
velocity difference between the colliding particles eventually becomes 
independent of the magnitude of the initial velocity difference.
Clearly, in the Boltzmann equation analysis this is not quite the case
and the peaks have a finite width as they converge.  However, this width
decreases in time as $1/t$, approaching the behavior of the
two-particle ring model asymptotically.

\section{Maxwell Problem}
\label{sec:Maxwell}
In the Boltzmann equation, the rate of collision of a pair of particles
depends on their relative velocity.  The ``Maxwell problem'' further
assumes that the pair of colliding grains is chosen randomly, all
pairs colliding with the same probability at unit rate.  The evolution
equation (\ref{Beq}) for the distribution of velocities is then replaced by
the even simpler form 
\begin{eqnarray}
\frac{\partial}{\partial t} P(v,t) &=& \int \int du_1\; du_2
P(u_1,t)P(u_2,t)\nonumber\\ \nonumber\\
&\times&\left[ \delta(v-u_1^\prime) + \delta(v-u_2^\prime)\right.
\nonumber\\ \nonumber\\
&&\left. -\delta(v-u_1) - \delta(v-u_2)\right].
\label{Meq}
\end{eqnarray}

\begin{figure}
\begin{center}
\includegraphics[width=8cm]{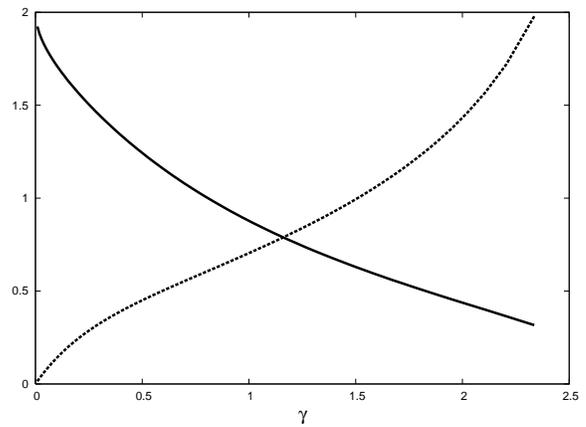}
\caption{Exponential decay parameter $\alpha$ 
(dashed curve), and algebraic decay parameter $\nu$ for the tails of the
distribution (solid curve), both as a function of viscosity $\gamma$.
\label{fig:maxwell}}
\end{center}
\end{figure}

This problem with a completely general linear collision rule (of which 
Eq.~(\ref{eq:vel}) is a special case) was analyzed in~\cite{our3}.
In the language used here, if we assume a scaling form as in
Eq.~(\ref{eq:scaling}), we arrive at an equation parallel to
Eq.~(\ref{Beq}) but now with the time-dependent factor
$-\dot{\phi}(t)/\phi(t)$.  Again, we argue that this term must be a
constant, and integration leads to an exponential decay of the width of
the distribution instead of the power law found in the Boltzmann case,
\begin{equation}
\phi(t) \sim e^{-\alpha t}.
\end{equation}
The decay constant depends on the friction coefficient $\gamma$.
In~\cite{our3} we described in detail how to evaluate $\alpha$, and in
Fig.~\ref{fig:maxwell} we show the resulting values of $\alpha$.
Furthermore, in~\cite{our3} we also analyzed the shape of the distribution.  We found
that (regardless of the initial distribution provided that
the initial
average velocity is zero) $F(x)$ has a single peak around $x=0$ instead
of the double-peaked
structure that we find for the Boltzmann problem and its two-ring
simplification, and that it has algebraic tails that decay as $F(x)\sim
x^{-\nu-1}$. The exponent $\nu$ is also a function of $\gamma$, as shown
in Fig.~\ref{fig:maxwell}.

\section{Numerical Simulations and Conclusions}
\label{sec:numerical}

The two approximations presented in the previous sections lead to very
different results for the velocity relaxation behavior of a
damped one-dimensional granular system.  One (Boltzmann problem) leads to a
velocity distribution that is double-peaked, with one at positive and one at
negative velocities, the peaks moving inward as $1/t$ toward a zero final
velocity.  The width of the peaks also decreases as $1/t$. 
The second model (Maxwell problem) leads to a distribution that is
single-peaked, with a width that decreases exponentially in time and
distribution tails that decay algebraically.  
Neither of these approximations considers the possible effects of the
spatial distribution of granules.  

\begin{figure}
\begin{center}
\includegraphics[width=8cm]{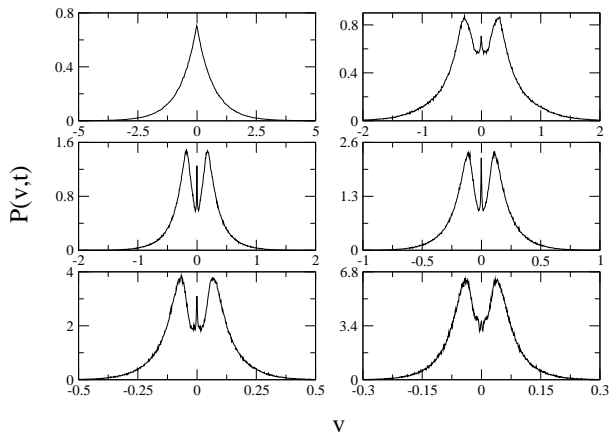}
\caption{
Velocity distribution obtained from numerical simulations on a line
with $10000$ granules averaged over $100$ realizations.  The particle
density is $0.001$, the initial spatial distribution is uniform, and the
initial velocity distribution is a symmetric exponential. The damping
parameter $\gamma=0.9$. From left to right and top to bottom, time is
1000, 8000, 16000, 32000, 64000 and 128000 in adimensional units. Notice 
the change in the scales as time proceeds. \label{fig:histogline}}
\end{center}
\end{figure}

\begin{figure}
\begin{center}
\vskip 10pt
\includegraphics[width=8cm]{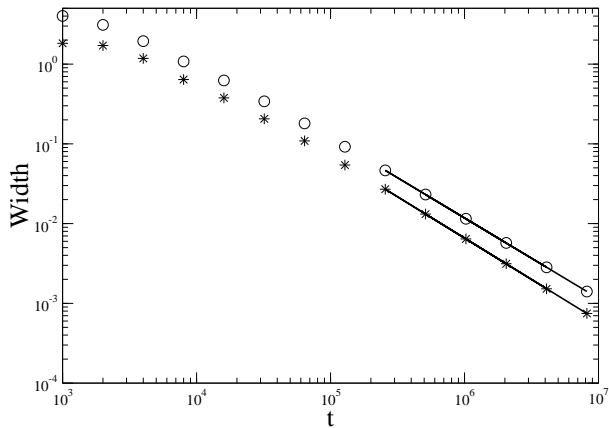}
\caption{Width of the velocity distribution (circles) and the 
width of each peak (stars). The lines represent the best fit
of the width vs time for the full distribution (exponent $-1.009$) and
for each peak (exponent $-1.035$). The initial velocity distribution
was a double Gaussian.\label{fig:widthline}}
\end{center}
\end{figure}

In order to assess the validity of these approximations, and also to get
a sense of the possible effects of the spatial distribution of granules,
we have carried out numerical simulations of a full chain of particles.
Our collision rules are still as indicated in Eq.~(\ref{eq:vel}), and, as
before, we assume collisions to be instantaneous, but now we actually
place the granules on a line and keep track of their positions so that
spatial inhomogeneities can occur if the system is so inclined.  Results
for $10000$ granules averaged over $100$ simulations are shown in
Fig.~\ref{fig:histogline}. 
The particle density is $10^{-3}$ with the particles
initially distributed uniformly.  The initial distribution is taken to
be a symmetric exponential, and $\gamma=0.9$.  
As time proceeds, the initial single peak splits into two peaks which
move inward and become
narrower, as predicted in our Boltzmann analysis.  Both the
inward motion of the peaks and the width of the peaks change
as $1/t$,  as in the Boltzmann approximation (see Fig.~\ref{fig:widthline}).

\begin{figure}
\begin{center}
\includegraphics[width=8cm]{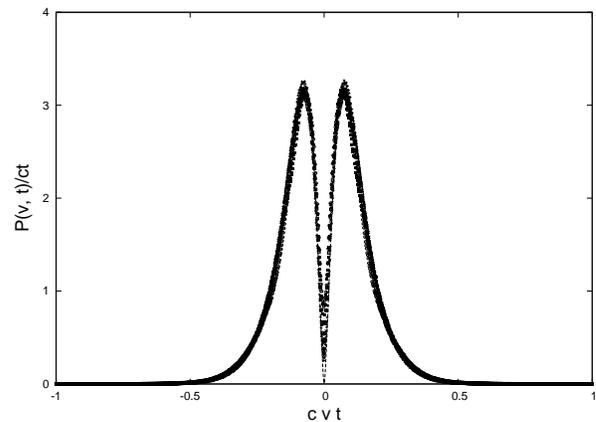}
\caption{Asymptotic behavior of the scaled velocity distribution. Different
symbols stand for different times
($1000\times 2^n$ with $n=7,8,\cdots,13$)
%(128000, 256000, 512000, 1024000, 2048000, 4096000 and 8192000)
and the dashed line is the best fit of a
double gamma distribution. The constant $c$ in the figure is 
arbitrary and was chosen to facilitate comparison with
Fig.~\ref{fig:scaling-boltz}.
\label{fig:scaling}}
\end{center}
\end{figure}

We also tested the scaling hypothesis Eq.~\ref{eq:scaling} on
our simulations. In Fig.~\ref{fig:scaling} we show that, as in the
Boltzmann approximation, the scaling works quite well. Moreover, 
in this case, the data can be fitted by a double $\Gamma$-distribution
\begin{equation}
F(x) = \frac{1}{2 b_1 \Gamma(b_2)} \left |\frac{x}{b_1}\right |^{b_2-1}
e^{-\left |\frac{x}{b_1}\right |}.
\end{equation} 
Note that this distribution has only two ($\gamma$-dependent) free
parameters, $b_1$, which sets the scale, and $b_2$, which sets the
shape. In Fig.~\ref{fig:scaling}
we plot this distribution for $\gamma=0.9$ ($b_1 = 0.046$ and $b_2 = 2.637$)
together with the simulation data obtained with a double Gaussian
initial velocity distribution.  We have tested the double
$\Gamma$-distribution fit for other values of $\gamma$, with equal success.

Comparing the velocity distribution for the Boltzmann
approximation (Figs.~\ref{fig:dist} and~\ref{fig:scaling-boltz})
and the simulations in a line (Figs.~\ref{fig:histogline}
and~\ref{fig:scaling}) we note that the Boltzmann model converges to the
scaling distribution much more rapidly. Initially the simulation
results exhibit the same behavior as the Boltzmann model, but for
longer times the behavior of the two distributions for small velocities
begin to differ.  In the Boltzmann case, the probability of finding
a slow granule is much greater than in the simulation. 

We conclude from these results that
\begin{enumerate}
\item
The Maxwell problem is an inadequate representation of the velocity
relaxation properties of the granular chain described in
Eq.~(\ref{eq:motion_rescaled}).  The predicted shape
of the distribution does not agree with that of the physical
chain.
\item
The Boltzmann problem in which we have disregarded spatial dependences
captures many of the essential features of the velocity relaxation, the
most important and new feature being the appearance of two peaks in the
velocity distribution.  The Boltzmann problem and even the simpler
two-particle simplification of the problem also capture the time
dependence of convergence of the two peaks into a single one at zero
velocity.  The slow ($1/t$) convergence is due to the fact that the
collision rate in the Boltzmann picture slows down as the gas cools.  By
way of contrast, the collision rate in the Maxwell problem does not,
leading to an exponential relaxation in time. 
\item 
The Boltzmann approximation does not correctly capture the
late time distribution of the slowest particles. 
This is probably due to spatial correlations that have been ignored in
this approximation and that are currently under
investigation~\cite{future}.
\end{enumerate}

One-dimensional momentum-conserving granular gases may exhibit
``inelastic
collapse,''  whereby the energy of the gas goes to zero in a finite
time~\cite{young1}.  Whether or not this occurs depends on the number of
particles $N$. There is a monotonically increasing relation between
$N$ and the coefficient of restitution $r$ for the critical value of
$N_c(r)$ above which collapse occurs, with $N_c\to\infty$ as $r\to 1$.
The
``quasi-elastic limit'' is the limit $r\to 1$ and $N\to \infty$, but in
such a way as to always remain below the collapse threshold.  In this
limit, a double-peaked velocity distribution is also
observed~\cite{young1,plus}.  In the
presence of friction there is no inelastic collapse, and we always
observe a double-peaked distribution.  A comparison
between those results and ours requires an understanding of spatial
correlations. This analysis will be presented elsewhere~\cite{future}.

Finally, we point out again that even in our most complete simulations we have
approximated the collision events as instantaneous.  While we do not
believe this to be a perceptible source of error, it would be
interesting (but extremely time consuming) to carry our full simulations
of the model Eq.~(\ref{eq:motion_rescaled}) with no further
approximations.

\section*{Acknowledgments}
This work was supported by the Engineering Research Program of
the Office of Basic Energy Sciences at the U. S. Department of Energy
under Grant No. DE-FG03-86ER13606 (KL) and by NSF Grant No. PHY-0140094 (DbA).

\end{document}